# Single Wavelength Operating Neuromorphic Device Based on a Graphene−Ferroelectric Transistor

K. Maity,[1] J.-F. Dayen,[1] B. Doudin,[1] R. Gumeniuk,[2] and B. Kundys[*,1]

[1] Université de Strasbourg, CNRS, Institut de Physique et Chimie des Matériaux de Strasbourg, UMR 7504, Strasbourg F-67000, France

[2] Institut für Experimentelle Physik, TU Bergakademie Freiberg, Freiberg 09596, Germany

**ABSTRACT:** As global data generation continues to rise, there is an increasing demand for revolutionary in-memory computing methodologies and efficient machine learning solutions. Despite recent progress in electrical and electro-optical simulations of machine learning devices, the all-optical nonthermal function remains challenging, with single wavelength operation still elusive. Here we report on an optical and monochromatic way of neuromorphic signal processing for brain-inspired functions, eliminating the need for electrical pulses. Multilevel synaptic potentiation−depression cycles are successfully achieved optically by leveraging photovoltaic charge generation and polarization within the photoferroelectric substrate interfaced with the graphene sensor. Furthermore, the demonstrated low-power prototype device is able to reproduce exact signal profile of brain tissues yet with more than 2 orders of magnitude faster response. The reported properties should trigger all-optical and low power artificial neuromorphic development based on photoferroelectric structures.

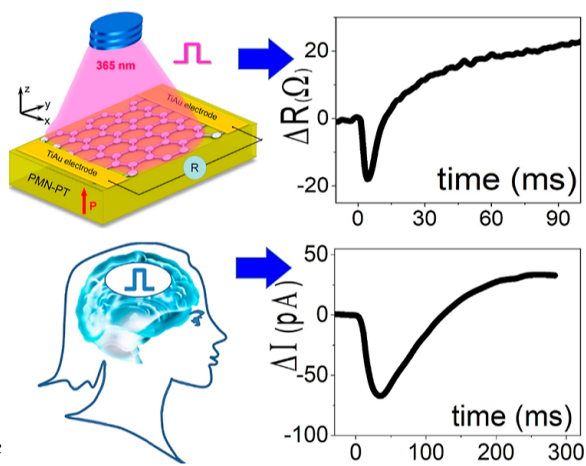

## ■ INTRODUCTION

Human brain inspires researchers for decades[1,2] and is now considered as an efficient way to bypass the separation between processing and memory units within the Von Neumann architecture of modern computers.[3] As transistors continue to shrink toward their physical limits,[4] the energy consumption and related heat are becoming major issues. Next generation computing technology architectures must be found for storing and processing outsized amounts of data[5] in a small space. The envisioned new approach must be inexpensive and able to provide efficient solutions for machine learning, enabling computers to perform tasks requiring human intelligence. Simulating human brain functions by electronics, therefore, is regarded as a promising way for creating more efficient and powerful computing systems.[6−10] Among functional solid state devices, those capable to demonstrate resistive switching can prone to energy and space-efficient memories potentially compatible with in-memory computing.[11] Here a special attention is paid to single-atom layer structures[12] which demonstrated extraordinary electronic properties providing unique opportunities for innovative devices control. In particular, the high sensitivity of graphene resistance to the presence of nearby electric charges makes graphene hybrid structures of great interest.[13] The powerful way to modulate charge environment is to use ferroelectric (FE) materials offering a large doping level modification and even polarity sign change, yet with the nonvolatile option to imprint multiple memory states.[14,15] Such structures can function with resistive readout of the ferroelectrically induced state, making device operation comparable with the resistive random access memories (RRAMs, memristors).[16] In electrically gated FE devices, graphene conductivity can be modified by several hundred of percents largely magnifying even tiny changes in charge environment. Furthermore, large changes in charge density in FE materials can be also induced optically[17,18] due to direct or photovoltaic (PV)-mediated photopolarization[19] Therefore, photoelectric (PE) properties of FE materials recently attract a renewed attention,[19,20] notably with the growing interest of PV[21] and neuromorphic applications.[22−25] Although recent reports have been successful in electro-optical control[26,27] for synaptic function including FE,[28−30] the entirely optical control yet at single-wavelength remains challenging. Here we report that such entirely optical control can be realized using PV free charge generation in FE. In electrical control using free charges in addition to dipoles can help to increase performance in FE electrically controlled transistors[31] and tunnel junctions.[32] However, static free charges lead to increased conductivity, and their density cannot be changed during device operation. In this regard, we

*kundysATipcms.fr

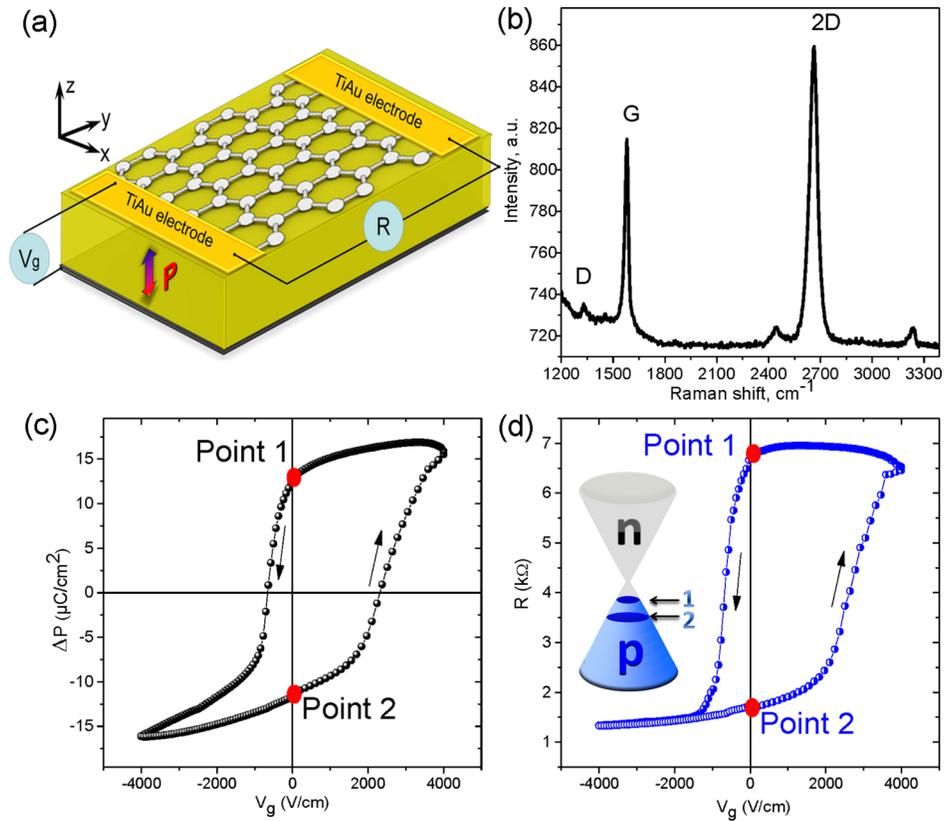

**Figure 1.** Device initial properties. (a) Schematics of the electrical experiment design; (b) Raman spectrum of the transferred CVD graphene layer over the (001) oriented $Pb[(Mg_{1/3}Nb_{2/3})_{0.70}Ti_{0.30}]O_3$ crystal. (c) FE loop of the substrate crystal; (d) related resistive hysteresis by the graphene layer measured in situ with the FE loop. Inset to the Figure (d) illustrates remanent doping levels with respect to the Dirac point.

propose to exploit the PV effect advantage in FE for deterministic charge generation. The both potentiation and depression cycles can be achieved fully optically yet with a single wavelength by using light fluence as a control parameter. The low-power coherent-less optical synaptic functions are attained via the PE effect for polarizing or depolarizing the structure on-demand while maintaining all other advantages of FE structures for artificial neuromorphics. Moreover, it is shown that different photoresponses can be achieved within the confines of a single cell, even when subjected to identical light excitation, demonstrating significant potential for enhancing the compactness and efficiency for neuron circuitry integration.

## ■ RESULTS

**Initial Electric State Characterization.** The as-prepared state of the structure was first characterized electrically to verify ferroelectricity and the related resistive response of the graphene before the optical excitations (Figure 1).

A clean graphene layer was obtained after chemical vapor deposition (CVD) as confirmed by the Raman spectrum (Figure 1b) as well as by the anticlockwise behavior opposite to FE/graphene structures with an additional interfacial dipole layer.[33−36] The clear resistive hysteresis of graphene that mimics the FE polarization loop indicates that graphene is intrinsically strongly doped (Figure 1d). Increasing the gate voltage in the positive direction increases the resistance between source and drain electrodes, therefore shifting the Fermi level toward the graphene Dirac point indicating the hole-dominated charge transport in graphene as illustrated by the inset to Figure 1d. The system does not cross the Dirac point as no peak in graphene resistance is observed by varying gate voltage in either direction. With decreasing the gate voltage, negative charges populate the FE surface and shift the Fermi level away from the Dirac point, rendering graphene more conductive for the negative polarization state [Figure 1d (inset)]. The trapping of electrons by the FE substrate makes graphene generally hole-doped in agreement with ref 37 with larger holes density at the point 2 (lower resistance). The two opposite FE polarization states of the substrate (points 1 and 2 in Figure 1c) lead to the formation of two remanent states in graphene resistance with nearly 300% change (Figure 1d). The FE loop also manifests the FE bias shift with coercive fields of +2.3 and −0.6 kV/cm (Figure 1c) equally translated to the graphene resistance hysteresis (Figure 1d).

**Optical Neuromorphic Function.** As multiple different polarization states exist in FEs in the subcoercive region,[38] the electric ground state was defined and stabilized in a reproducible way prior to optical excitations. To this end, the structure was electrically poled in darkness to reach the remanent FE state noted as point 1 in Figure 1c,d. This was done by sweeping electric field from 0 to −4 kV/cm followed by −4 to +4 kV/cm sweep and then back to zero. Time dependent measurements of the electric polarization were then conducted to ensure polarization stability. In Figure 2a (upper panel), the electrically prepared device is exposed to a UV light pulse of 365 nm, of 3.4 eV exceeding the bandgap energy of the FE substrate, of 3.2 eV.[39] Depending on the light fluence (intensity and pulse duration), the graphene resistance response is drastically different. The long-term memory effect is observed at a light fluence of ∼0.2 mJ/cm$^2$ (38 mW/cm$^2$ pulse during 5 ms), as depicted in Figure 2a (upper panel). The signal profile closely resembles that observed in biological brain tissues.[40] Notably, our optical control demonstrates a significantly faster response, surpassing 2 orders of magnitude in comparison to bio samples, as illustrated in Figure 2a (lower panel). The signal depression is achievable for light fluencies as small as ∼0.4 μJ/cm$^2$ (74 μW/cm$^2$, 5 ms) and yields a decrease of resistance with the negative saturation trend (Figure 2b). It is therefore evident that there are two superposed mechanisms involved, which compete as a function of light fluence. The light-induced temperature increase in the sample was found negligible at most intensities employed, only becoming detectable for intensities >25 mW/cm$^2$ resulting in sample warming below 0.7 K with a linear response (Figure S1). In order to get insight into the origin of the observed effects, the light intensity dependent study was performed and is illustrated by Figure 2c. In agreement of the data in Figure 2a,b, graphene resistance is also decreasing for the light intensities below the threshold of 1.1 mW/cm$^2$ (100 mJ/cm$^2$) where it reaches a minimum. However, beyond this threshold, the resistance begins to increase, exceeding the initial resistance for higher light intensities. To explain this behavior, one should recall the depolarization–pyroelectric[41,42] and PV[21,43,44] competing effects in FEs. The contributions to electrical current across the sample therefore come from both dipoles and photocarriers if the FE is also PV. Indeed, the bulk source-gate current ($I_{sg}$) across the FE substrate shown on Figure 2c (right scale) matches the initial declining slope of the graphene sheet resistance. As the light intensity continues to increase, the $I_{sg}$ current also rises, indicating a progressively significant PV effect. However, if the excitation wavelength is of bandgap energy, the pyroelectric response also occurs on the background of photocarriers generation. Therefore, the photoresponse can vary with intensity and wavelength of the incident light. At lower intensities, the generation of free charges within the insulating FE begins to reduce the overall polarization. This continues until a sufficient number of photocarriers accumulate to compensate the dipoles polarization. Conversely, under yet higher light intensities, an excess of photocarriers causes them to recombine with pre-existing screening charges, ultimately leading to subsequent polarization recovery. We can therefore take advantage of these two competing processes termed here depolarization and polarization to realize the all-optical control for on-demand processing of the graphene resistance signal. More precisely, our initially hole-doped graphene layer becomes less or more positive, depending on the resting electrostatic equilibrium after the PV light excitation. In fact, biological neurons undergo similar polarization and depolarization processes when electrical pulses render ion channels to open making the neuron system more or less polarized due to charged ions flow.[45] Hence, we can exploit this similarity to achieve neuromorphic function. As an illustration of this strategy, a set of 20 pulses of the larger fluence was used to potentiate the signal, while same number of pulses, but of smaller fluence was used to depress it (Figure 2d). Moreover, the stable resistance level can be formed in-between after some relaxation, demonstrating a long-term memory effect. The relaxation can be attributed to Debye relaxation, which is intrinsic to the dielectric properties. In fact, a similar relaxation at the same time scale is also observed after depression. For further testing, the ten times larger pulse quantity (of 200) was successful and resulted in increased states resolution (Figure S2). The two different natures of mechanisms responsible for resistance increase and decrease are confirmed in the zoomed regions of the corresponding slopes (Figure S2).

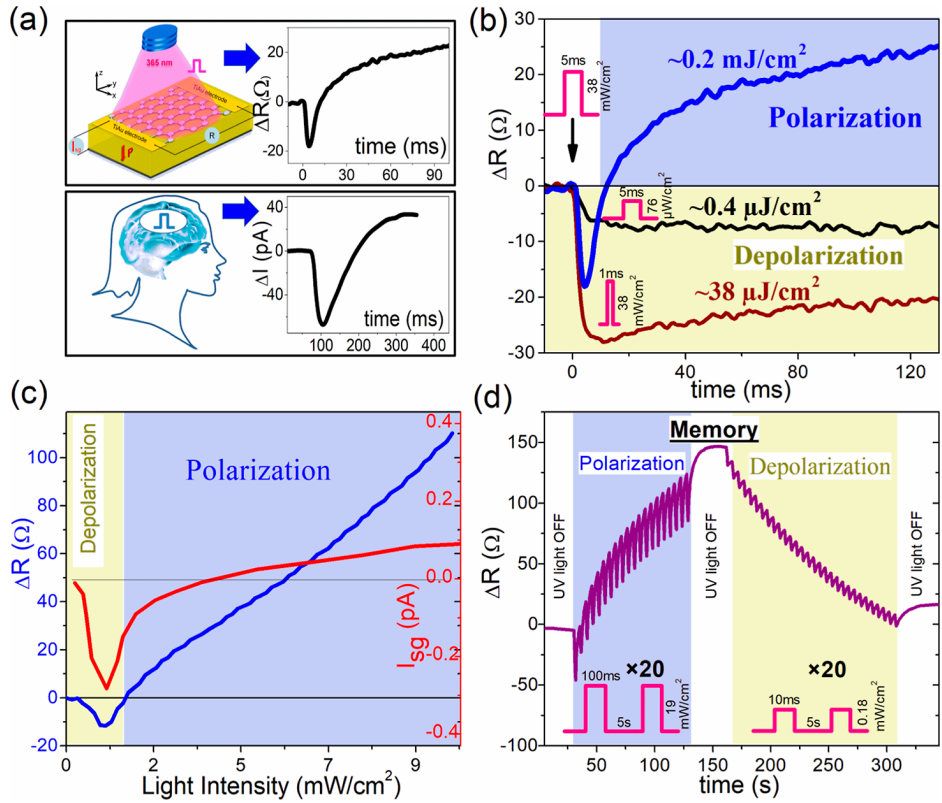

**Figure 2.** Artificial optical neuromorphic function. (a) Schematics for optical excitation (upper panel) with a similar brain neuron response reproduced with permission from ref 40 (lower panel). (b) Graphene resistance response to the UV light pulse of different fluences. (c) Light intensity dependence of the graphene resistance and related source-gate current. (d) All-optical single wavelength potentiation and depression of the graphene resistance.

## ■ DISCUSSION

The observed on-demand potentiation and depression of the graphene resistance can be analyzed via the qualitative model, as shown in Figure 3. In the initial state, the sample is polarized via dipoles 180° rotation (remanent polarization state 1) with internal depolarization field and no free charges in the ideal case (Figure 3a). When UV light is on, the temporal reduction of polarization can occur via pyroelectricity on the background of photocarriers generation. The photocarriers then distribute to electrodes increasing depolarization and rising graphene hole doping level in agreement with initial resistance decrease (Figure 2c). This process occurs until dipole polarization is fully screened by photocarriers. If light intensity is sufficient to create excess of photocarriers, they change their flow direction, starting to recombine. If light is still on, the PV current can stabilize (Figure 3c). When the light is off, a new electrostatic equilibrium is formed between the dipole polarization and the depolarization by free charges. Thus, repetitive light excitation can be used to either increase or decrease the graphene resistance, depending on the amount of generated photocarriers at the given light intensity and at the given ground state. Although the model can be more complicated,[46] the basic mechanism can be simplified here to the competition between FE polarization and depolarization by photocarriers. In this way, one can achieve the four basic processes of synaptic plasticity (Figure 4). Namely, long-term potentiation (LTP), long-term depression (LTD), short-term potentiation (STP), and short-term depression (STD), which occur in the hippocampus brain region responsible for storing memories. While generally smaller energy (smaller amount of photocarriers) is needed for the resistance level to be recovered after excitation (Figure 4a,b), the larger energies are required to induce the long-term remanent effects (Figure 4c,d), as illustrated in Figure 4e. Due to long-term memory effects, a deliberate decision was made to operate at a logic level which is recoverable through light exposure of corresponding pulses after each illumination. Here, FE offers the advantage of enabling electrical resets at any time simply by returning it to its remanent point 1 (Figure 1c,d) in the absence of light. To further get insight into the photocarriers contribution, the photoresponse was measured also for below bandgap excitation energies i.e., 530 and 940 nm wavelengths (Figure 5a). Because no potentiating was possible for the 530 and 940 nm wavelengths, the same UV pulse (365 nm of 100 ms duration) was used to increase the resistance prior to depression function testing. Depression functions are also attainable at 530 and 940 nm wavelengths (Figure 5a), and they can be used for inducing a logic level shift, enhancing neuromorphic plasticity. The absence of potentiation for the 530 and 940 nm wavelengths indicates their inability to generate free charges inside the FE by band-to-band transition. Indeed, there is no PV current for 530 and 940 nm wavelengths (Figure 5b); however, the pyroelectric effects still exist. Therefore, it can be concluded that ability to increase the graphene resistance in our device is due to PV current generation, while the ability to depress resistance by the below bandgap wavelengths is likely due to pyroelectric depolarization possibly involving interfacial trap states between graphene and FE.[30,47] However, the pyroelectric response at 365 nm was found to be faster and more efficient (a smaller light fluence required to depress) compared to the one at 530 nm light as it yields simultaneous photocarriers generation (Figure S3). Therefore, in the single-wavelength operation with above bandgap light energy, our depolarization used to depress signal consists of the 2-fold response of FE dipoles and photocarriers. The last one is much faster and can be used to create reconfigurable neuromorphic-like system patterns. Moreover, taking into account that the PV effect in FE can be nonsymmetric with respect to poling sign[38] (i.e., different PV responses at remanent FE states 1 and 2 due FE bias and stresses), one can program the multicell behavior, as shown in Figure 6. The light pulse of the same energy can depress one artificial synapse while potentiating another at the given FE ground state, enabling a higher density of neural integration.

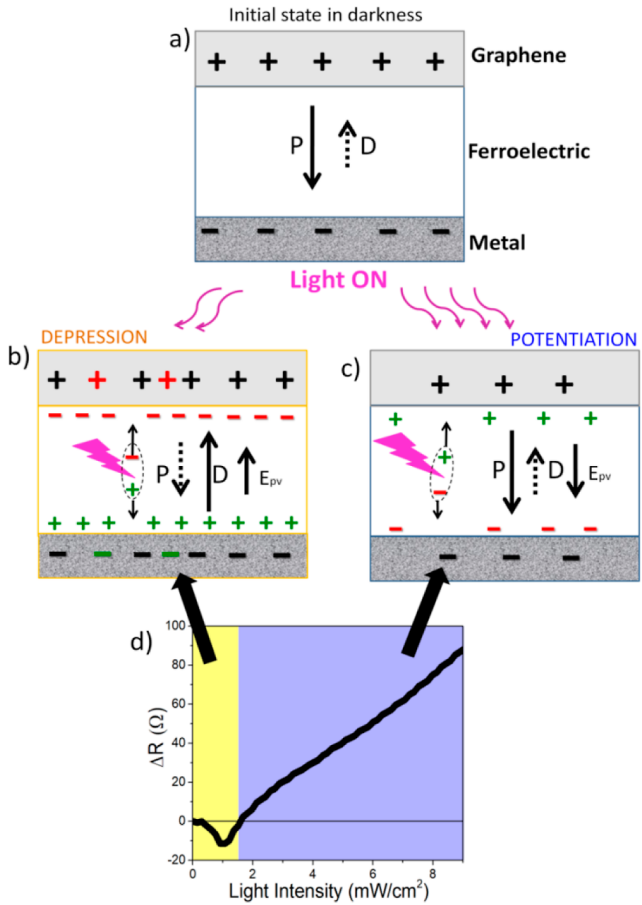

**Figure 3.** Model for photogenerated charge carrier dynamics. The initial polarization (a) is screened by photocarriers (b). If light intensity is above the threshold, recombination occurs, and PV current stabilizes (c). When light is off, both above and below threshold excitations lead to different electrostatic states and related graphene doping levels reflecting themselves in the decreased or increased graphene resistance (d).

## ■ CONCLUSIONS

In summary, an entirely optical monochromatic way to potentiate and depress graphene resistance was demonstrated in the straightforward two component graphene/FE memristor device. The background mechanism is a competition between the polarization and depolarization processes governed by the amount of photogenerated charges used to modify graphene overlayer doping level. Because similar polarization–depolarization mechanisms are central in biological neurons, our approach successfully emulates basic brain functions with a wireless excitation advantage at a single wavelength yet on much faster time scales than in biological systems. Both short- and long-term memories with a large number of intermediate states can be created thanks to a nonlinear response with light intensity. While the ability to potentiate relates to the intrinsic PV effect specific to close to bandgap wavelengths, the depression effect can be achieved even for larger wavelengths via pyroelectric depolarization, although at slower response.

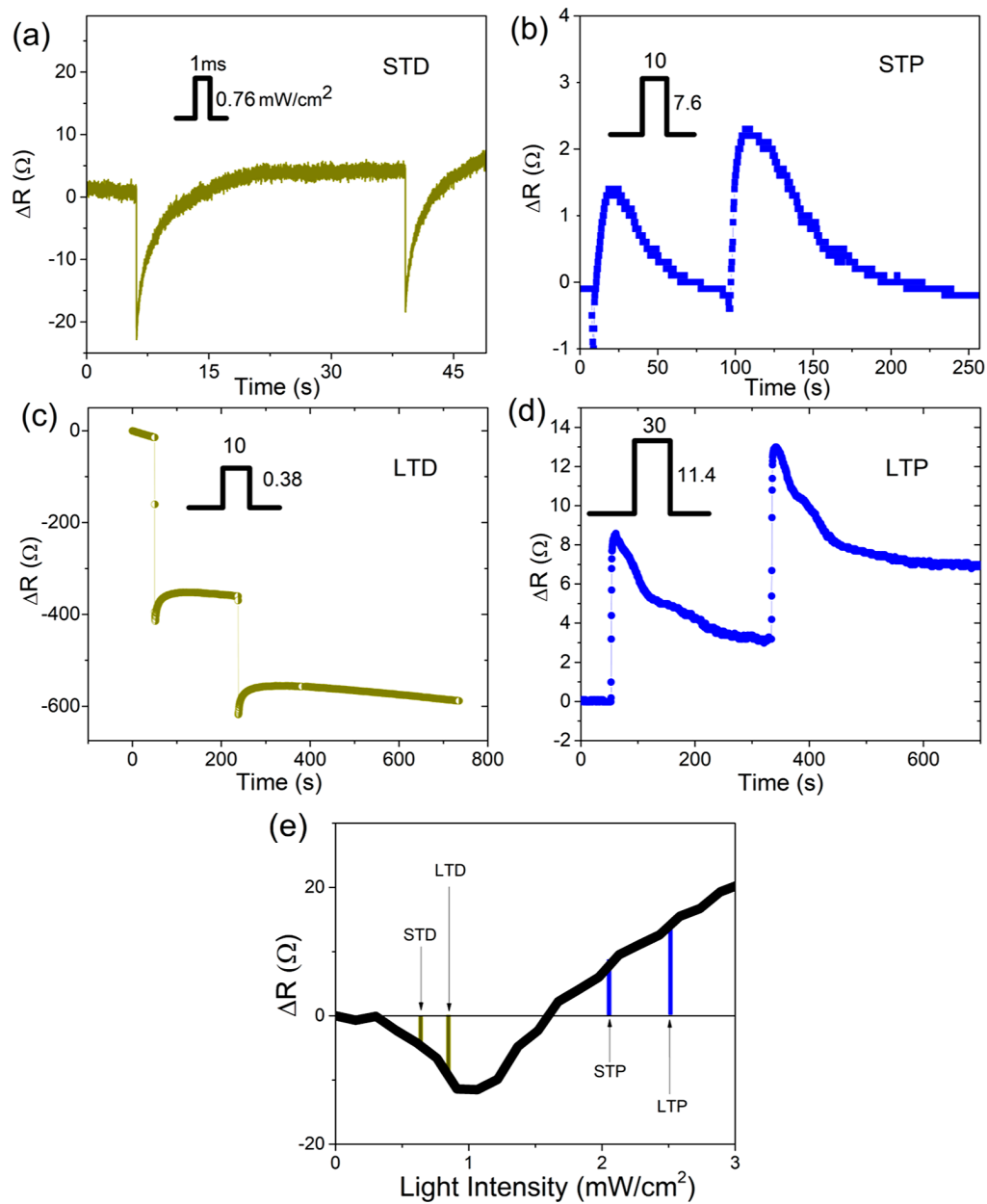

**Figure 4.** Neuromorphic signal processing. The four basic functions: STD (a), STP (b), LTD (c), and LTP (d). Figure (e) illustrates the intensity conditions for reversible and irreversible switching.

The below comparison Table 1 summarizes up-to-date-optical methods[48−57] to simulate neuromorphic function in different structures. While the reported here PV–FE method remains competitive in terms of light intensities, it excels in single-wavelength operation and structure simplicity with a potential to enhance the neural integration density in polydomain FE structures (Figure 6).

More generally, the effects of extrinsic photocarriers on transport properties of graphene in the straightforward two component device are extendable to other 2D-based hybrid FE systems and should help to analyze more complicated hybrid–FE interfaces. From the perspective point of view, the all-optical switching of graphene's conductivity is highly desired for integrated photonics[58] and plasmonics,[59] where graphene plays a role of the waveguide for high-speed operation. Therefore, realization of a photogated graphene waveguides is a promising research undertaken for scalable photonic platforms based on all-optical photoferroelectric devices[60,61] possibly extendable even to quantum-technology applica-tions.[62] In a short-term perspective, this study should serve for a better understanding of the light-induced FE polarization modifications in FEs, paving the way for optimized low-power and all-optical function in the artificial neuron systems.

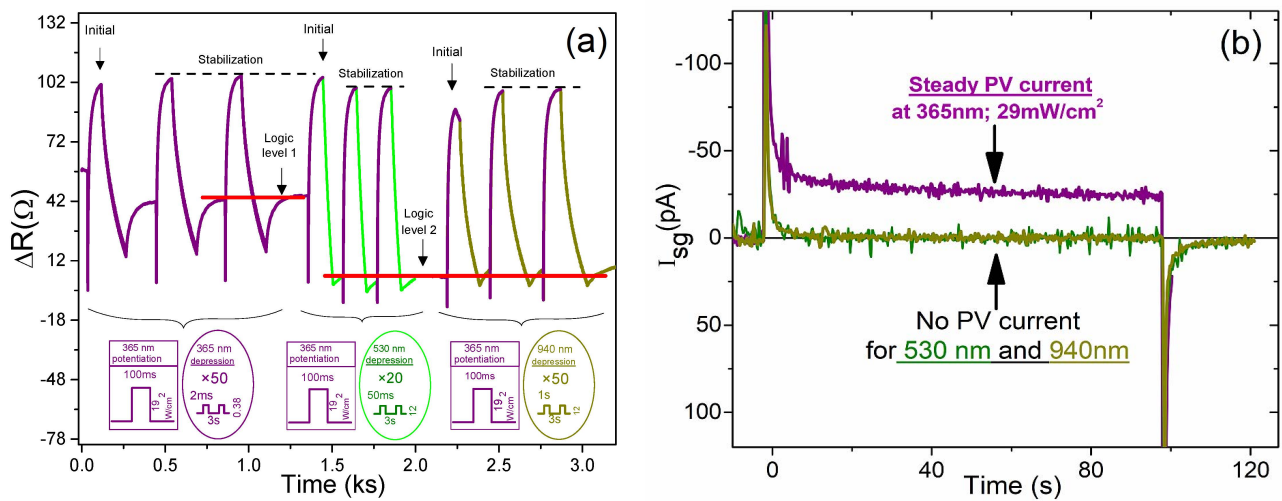

**Figure 5.** Optical processing of the graphene resistance signal. The same UV pulse was used for potentiation, while depression was done by 365, 530, and 940 nm (a). The disability of optical potentiation for the 530 and 940 nm wavelengths is explained by the absence of PV current (b).

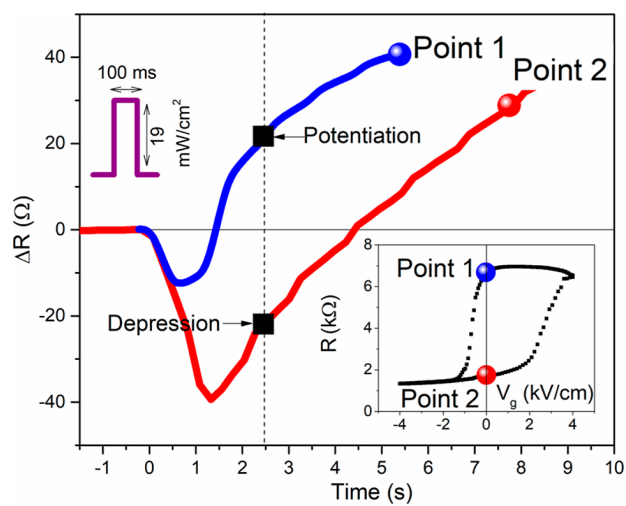

**Figure 6.** Remanent FE state-dependent optical response. Same intensity of 356 nm light excitation can produce different responses for different FE states (inset) due to PV effect asymmetry in FEs.

## ■ MATERIALS AND METHODS

The single crystals of $Pb[(Mg_{1/3}Nb_{2/3})_{0.70}Ti_{0.30}]O_3$ were ordered from Crystal GmbH (Germany). These compounds are low-electric field switchable FE crystals of so-called PMN-PT family reported to manifest large scale of interesting properties ranging from piezoelectric[63,64] and electro-optic[65,66] to PV[67,68] and photostrictive.[69,70] After CVD graphene deposition, the sample was annealed and a piece of 1.87 mm by 0.82 mm was cut from the 10 mm² wafer of 0.3 mm thickness. The lateral drain and source electrodes deposition of Ti(5 nm)/Au(30 nm) layers was done by E-beam evaporation via a shadow mask and wired using the silver conductive paste. The FE loop was measured along the shortest sample dimension ([001]) using a homemade quasi-static FE at 5 mHz. The overall CVD graphene preparation procedure was found to reduce the remanent FE polarization and induce a FE bias shift toward positive electric fields (Figure 1c). We used above bandgap excitation[39] provided by a light emitting diode (LED) model M365FP1 (Thorlabs) at 356 and 9 nm bandwidth calibrated with a Thorlabs power meter (model PM100USB). Before optical excitation, the sample was electrically poled in darkness by sweeping the electric field from zero to −4 kV/cm followed by −4 to +4 kV/cm sweep and then back to zero (remanent point 1). The poling of the inverse sign was used to obtain remanent point 2. The light intensity-dependent study was done by sweeping light intensity from 0 to 10 mW/cm$^2$ with a step of 0.26 mW/cm$^2$.

**Table 1. Non-Electrical Optical Neuromorphic Methods for Different Structures Reported**

| sample | wavelengths (nm) | minimal intensities (mW/cm²) | mechanism | refs |
|---|---|---|---|---|
| PtTe$_2$/Al$_2$O$_3$/MoS$_2$ | multi 300 nm−2 μm | 0.22−0.36 | trapping/detrapping of photocarriers | 48 |
| layered black phosphorus | multi 280, 365 | 3−6.8 | trapping/detrapping of photocarriers | 49 |
| copper-phthalocyanine (CuPc) and (P(VDF-TrFE)) | multi 660, 445 | 28−56 | trapping/detrapping of photocarriers | 50 |
| CH$_3$NH$_3$PbBr$_3$−ZnO$_2$ heterostructure | multi 365, 460, 420 | 8.75−32.5 | ionization−neutralization of oxygen vacancies | 51 |
| hBN/WSe$_2$ | multi 655, 532, 405 | 6 | trapping/detrapping of photocarriers | 52 |
| Pyr-GDY/graphene/PbS-QD | multi 635, 365 | 0.6−1.2 | photoinduced oxygen-desorption | 53 |
| a-IGZO/CdS 3T | multi 620, 525 | 0.98−0.092 | trapping/detrapping of photocarriers | 54 |
| ZnO/PbS 2T | multi 365, 980 | | ionization−neutralization of oxygen vacancies | 55 |
| Ag/TiO$_2$T | multi 632, 360 | 142.9−1.43 | oxidation/reduction processes of the Ag nanoparticles | 56 |
| PEA2SnI4/Y6 3T | multi 808, 450, 520, 650 | 0.025 | trapping/detrapping of photocarriers | 57 |
| **graphene/PMN-PT** | single 365 | 0.76−11 | photovoltaic−ferroelectric | this work |

## ● Supporting Information

The Supporting Information is available free of charge at https://pubs.acs.org/doi/10.1021/acsami.3c10010 and below.

The Supporting Information details on sample warming evaluation, potentiation−depression cycles with increased number of pulses and depression time profile comparison for 365 and 530 nm wavelengths.

## Author Contributions

B.K. and K.M. initiated the ex periment. K.M. conducted measurements under supervision with participation of B.K. The funding was accrued by B.K., B.D., and J.F.D. All authors contributed to the analysis of the results and scientific discussion. The manuscript was written by B.K. through contributions of all authors.

The authors declare no competing financial interest.

## ■ ACKNOWLEDGMENTS

We acknowledge the financial support from IdEx 2022—AAP Recherche Exploratoire program of Strasbourg university, ANR MixDFerro grant (ANR-21-CE09-0029), and Institut Universitaire de France. We also thank F. Chevrier and N. Beyer for technical help.

## ■ REFERENCES

**Supporting Information**

**Single wavelength operating neuromorphic device based on graphene-ferroelectric transistor**

Krishna Maity[1], Jean-François Dayen[1], Bernard Doudin[1], Roman Gumeniuk[2], Bohdan Kundys*[1]

[1]Université de Strasbourg, CNRS, Institut de Physique et Chimie des Matériaux de Strasbourg, UMR 7504, 23 rue du Loess, Strasbourg, F-67000, France.

[2] Institut für Experimentelle Physik, TU Bergakademie Freiberg, Leipziger Str. 23, Freiberg 09596, Germany.

*To the Correspondence

Name: Bohdan Kundys

* E-mail: Kundys[a]ipcms.fr

Mailing address: Institut de physique et de chimie des matériaux de Strasbourg (IPCMS)

23 Rue du Loess, Bâtiment 69, 67200 Strasbourg, France.

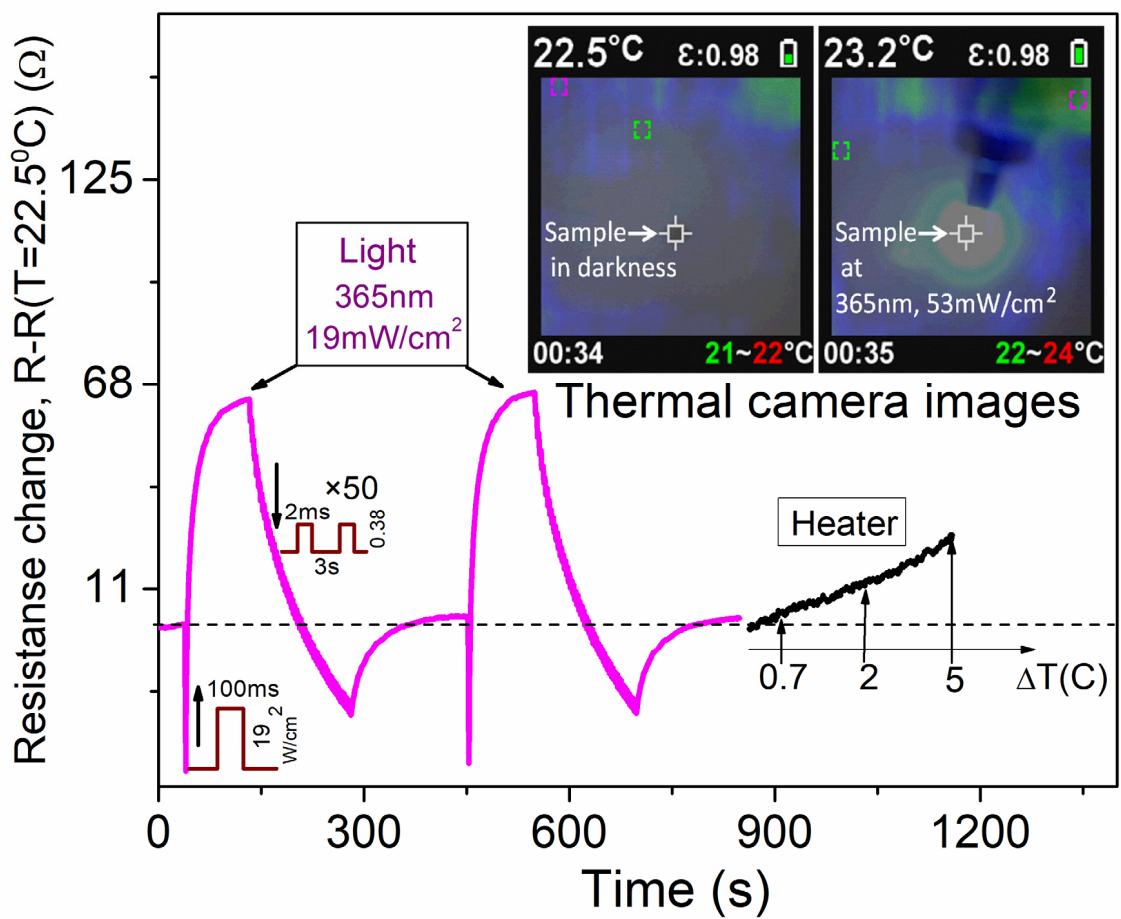

**Figure S1.** UV light-induced sample warming assessment. Thermal camera images (insets) evidence 0.7$^0$C warming under 53mW/cm$^2$ of 365nm light intensity that is negligible with respect to photo-ferroelectric effect induced by 19 and 0.38mW/cm$^2$ light intensities.

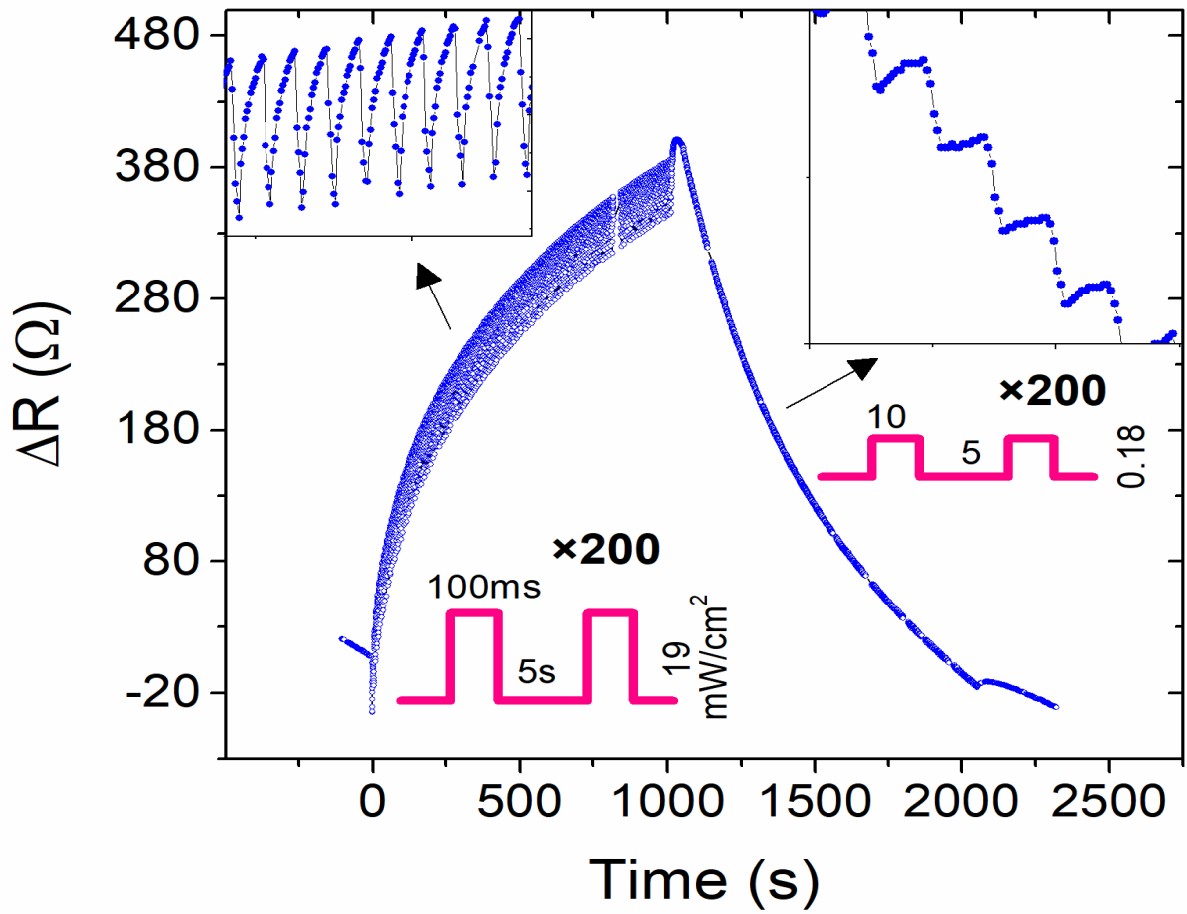

**Figure S2.** Optical potentiation and depression cycle with increased number of pulses. The insets represent zoomed regions of potentiation and depression.

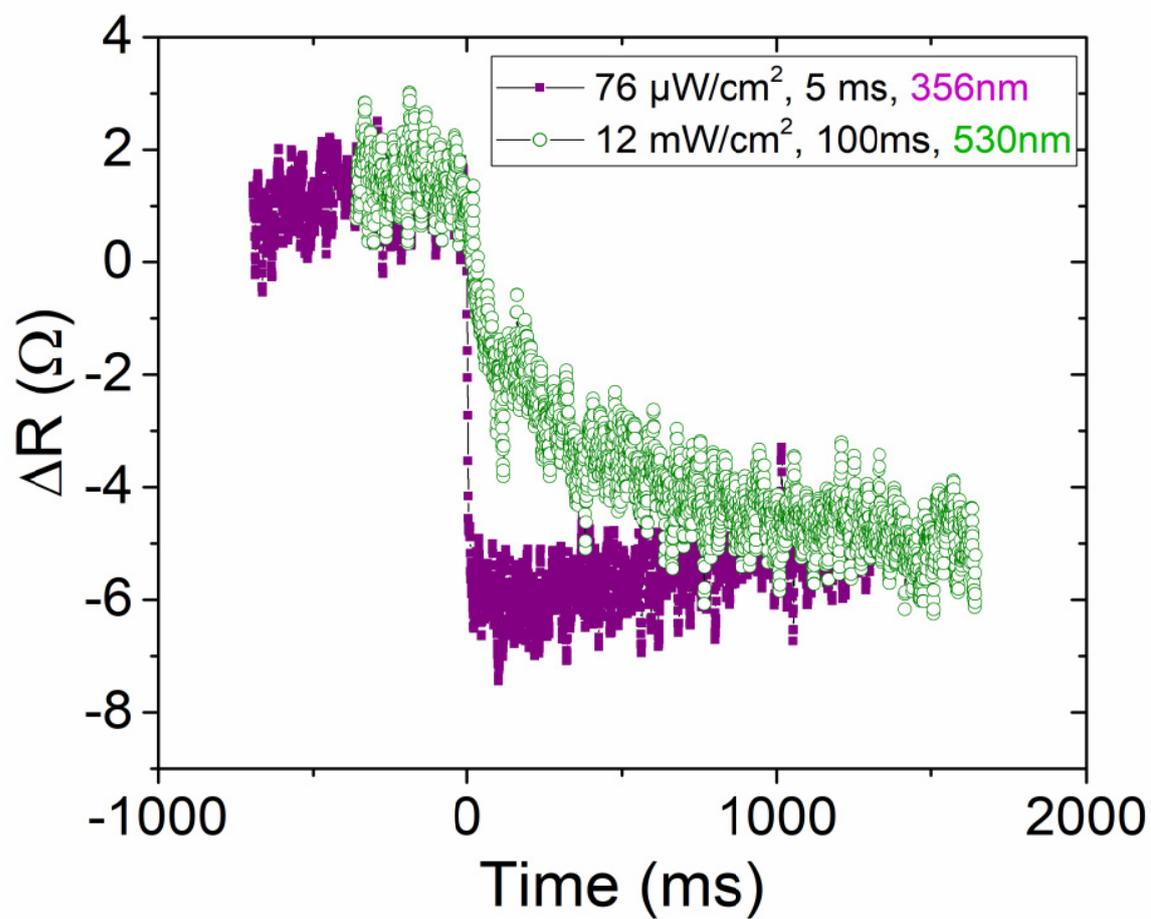

**Figure S3.** Depression time profile for green and ultraviolet light. Due to photocarriers the 365nm wavelength leads to faster response compared to the green (530nm) wavelength.